\documentclass[aps,prb,twocolumn,amsmath,amssymb,showpacs,superscriptaddress]{revtex4-1}

\usepackage{graphicx}
\usepackage{amsmath,amssymb,amsfonts}
\usepackage{textcomp}
\usepackage{hyperref}
\usepackage{gensymb}
\usepackage{verbatim}
\usepackage{amsmath}
\hypersetup{breaklinks=true,colorlinks=true,urlcolor=black}
\usepackage{color}
\usepackage{soul}
\DeclareGraphicsExtensions{.jpg,.pdf,.png,.eps}

\usepackage{multirow}
\usepackage{dcolumn}

\begin{document}
\title{Optimization of the Crystal Growth of the Superconductor CaKFe$_{4}$As$_{4}$ from Solution in the FeAs-CaFe$_{2}$As$_{2}$-KFe$_{2}$As$_{2}$ System}
\author{W. R. Meier}
\affiliation{Ames Laboratory US DOE and Department of Physics and Astronomy, Iowa State University, Ames, Iowa 50011, USA}
\author{T. Kong*}
\affiliation{Ames Laboratory US DOE and Department of Physics and Astronomy, Iowa State University, Ames, Iowa 50011, USA}
\author{S. L. Bud'ko}
\affiliation{Ames Laboratory US DOE and Department of Physics and Astronomy, Iowa State University, Ames, Iowa 50011, USA}
\author{P. C. Canfield}
\affiliation{Ames Laboratory US DOE and Department of Physics and Astronomy, Iowa State University, Ames, Iowa 50011, USA}
\email[]{canfield@ameslab.gov}

\date{\today}

\begin{abstract}
Measurements of the anisotropic properties of single crystals play a crucial role in probing the physics of new materials. Determining a growth protocol that yields suitable high-quality single crystals can be particularly challenging for multi-component compounds. Here we present a case study of how we refined a procedure to grow single crystals of CaKFe$_{4}$As$_{4}$ from a high temperature, quaternary liquid solution rich in iron and arsenic ("FeAs self-flux"). Temperature dependent resistance and magnetization measurements are emphasized, in addition to the x-ray diffraction, to detect inter-grown CaKFe$_{4}$As$_{4}$, CaFe$_{2}$As$_{2}$ and KFe$_{2}$As$_{2}$ within, what appear to be, single crystals. Guided by the rules of phase equilibria and these data, we adjusted growth parameters to suppress formation of the impurity phases. The resulting optimized procedure yielded phase-pure single crystals of CaKFe$_{4}$As$_{4}$. This optimization process offers insight into the growth of quaternary compounds and a glimpse of the four-component phase diagram in the pseudo-ternary FeAs-CaFe$_{2}$As$_{2}$-KFe$_{2}$As$_{2}$ system.

\end{abstract}

\pacs{}
\maketitle 

\section{INTRODUCTION}
Single crystals of the iron arsenide superconductors have been grown by a variety of techniques. The parent compounds of the 122 family CaFe$_{2}$As$_{2}$ \cite{Goldman2009_CaFe2As2Collapse_NeutronXray,Ran2012CaFe2-2xCo2x2As2_Annealing}, SrFe$_{2}$As$_{2}$ \cite{Matsubayashi2009AFe2As2Pressure}, and BaFe$_{2}$As$_{2}$ \cite{Matsubayashi2009AFe2As2Pressure,Wang2009BaFe2As2}, can all be grown as single crystals out of solutions rich in iron and arsenic. For each of these compounds, superconductivity can be stabilized by partially substituting the alkaline earth cation with a similarly sized alkali metal ion.\cite{Rotter2008Ba1-xKxFe2As2Superconductivity,Ni2008Ba1-xKxFe2As2,Wang2008Ba1-xKxFe2As2Hc2,Bukowski2009Ba1-xRbxFe2As2Superconductivity,Gen-Fu2008Sr1-xKxFe2As2Superconductivity,Shirage2008Ca1-xNaxFe2As2} For example, (Ba$_{0.6}$K$_{0.4}$)Fe$_{2}$As$_{2}$ is a superconductor with a transition temperature of 38\,K.\cite{Rotter2008Ba1-xKxFe2As2Superconductivity,Paglione2010_FeBasedSuperconductivity} Single crystals with these solid solutions can often be achieved without significant modification of the growth parameters for many of the compositions.\cite{Ni2008Ba1-xKxFe2As2,Liu2014Ba1-xKxFe2As2_Growth}

Recently, Iyo et al.\cite{Iyo2016AeAFe4As4} reported that for some pairs of alkali and alkaline earth elements of different ionic size, the cations segregate into different crystallographic sites. This leads to the stoichiometric \textit{AeA}Fe$_{4}$As$_{4}$ (\textit{Ae} = Ca, Sr and \textit{A} = K, Rb, Cs) compounds with alternating \textit{Ae} and \textit{A} ion layers along the crystallographic c-axis.\cite{Iyo2016AeAFe4As4} They also showed that all of the members of this family of compounds are superconducting with $T_{\mathrm{c}}$'s between 31 and 36\,K, as ordered line compounds without additional doping.

Solution growth of singe crystalline \textit{AeA}Fe$_{4}$As$_{4}$ is not only challenging because it entails growing a stoichiometric, four-element, compound from a 4-element melt. In addition, the target \textit{AeA}Fe$_{4}$As$_{4}$ (1144) phase and its parent 122 phases (\textit{Ae}Fe$_{2}$As$_{2}$ and \textit{A}Fe$_{2}$As$_{2}$) are all structurally and compositionally similar. In fact, Iyo et al. suggest that the similarity of the crystallographic a-lattice parameters between the two related 122's is necessary for formation of the 1144 phase.\cite{Iyo2016AeAFe4As4} As a result, these 122 phases compete with 1144 during crystallization and turn out to be persistent impurities. 

In this paper, we describe how we optimized the high temperature solution growth of single crystals of CaKFe$_{4}$As$_{4}$ out of an Fe-As rich solution. The anisotropic properties of these crystals are published in reference~\onlinecite{Meier2016CaKFe4As4}. In addition to x-ray diffraction, we relied on magnetization and resistance measurements to characterize growth products. Guided by these results, we adjusted growth parameters to suppress formation of the CaFe$_{2}$As$_{2}$ and KFe$_{2}$As$_{2}$ impurity phases which we observed to inter-grow with CaKFe$_{4}$As$_{4}$. The results of the many growths made during this process provide insight into the solidification behavior of liquids in this composition region. We use this information to propose a phase diagram for the pseudo-ternary FeAs-CaFe$_{2}$As$_{2}$-KFe$_{2}$As$_{2}$ system.

\section{EXPERIMENTAL METHODS}
\subsection{Growth Assembly and Details}

\begin{figure}
	\includegraphics[width=\columnwidth]{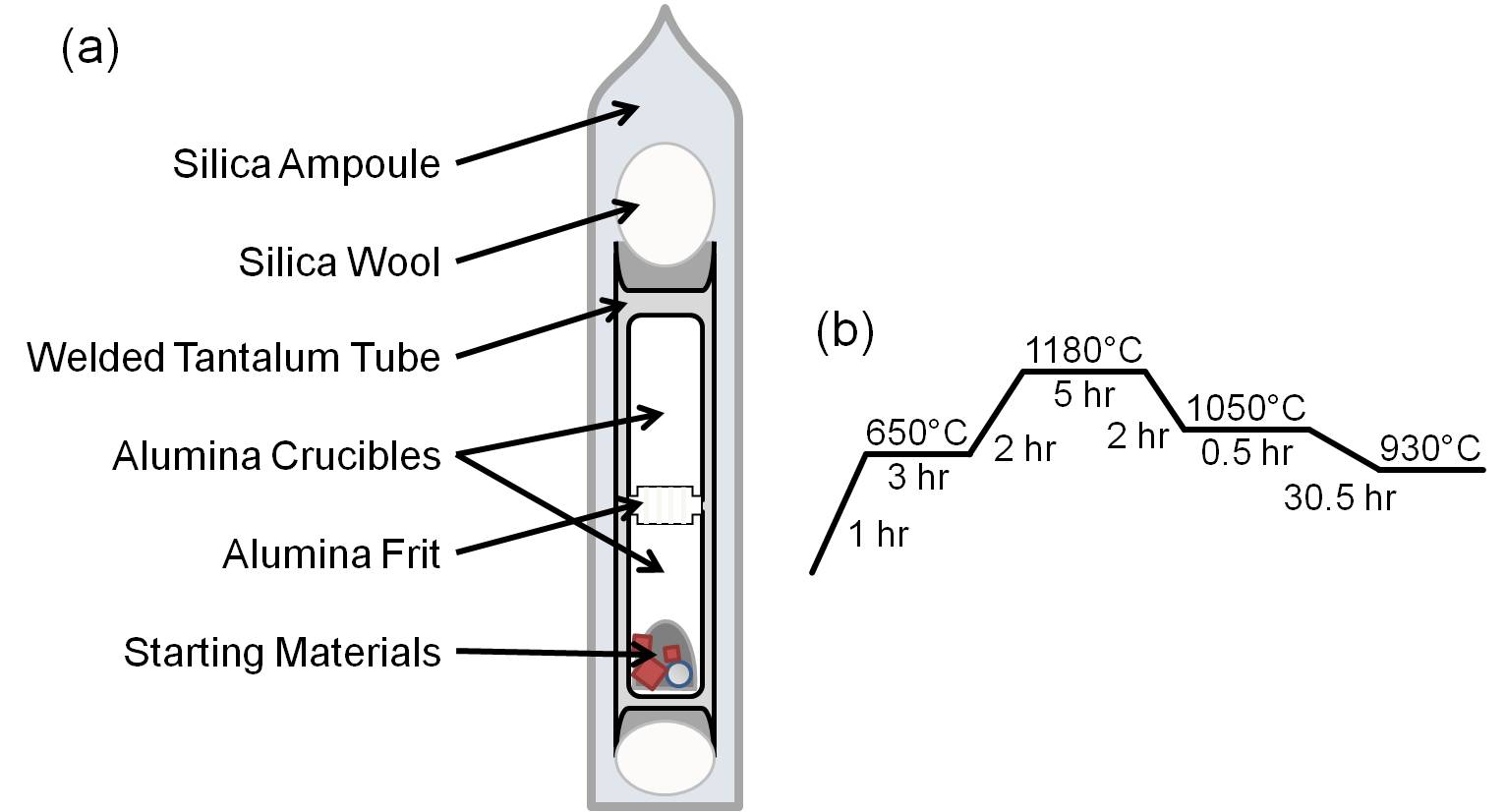}%
	\caption{(a) Ampoule assembly showing nesting of components and (b) a schematic of the optimized furnace schedule used in batch \textbf{F1}.
		\label{ampoule_profile}}
\end{figure}

All of the CaKFe$_{4}$As$_{4}$ growth attempts described here used high temperature solution growth employing liquids with excess iron and arsenic ("FeAs self-flux"). Lumps of potassium metal (Alfa Aesar 99.95\%), distilled calcium metal pieces (Ames Laboratory, Materials Preparation Center (MPC) 99.9\%) and Fe$_{0.512}$As$_{0.488}$ precursor powder were loaded into a 1.1\,x\,1.1\,x\,2.4\,cm$^{3}$ fritted alumina Canfield Crucible Set (LSP Industrial Ceramics, Inc.) in an argon filled glove-box.\cite{Canfield2016FritDiskCrucibles} Batches totaled about 2\,grams. The Fe-As precursor was synthesized by solid state reaction at 900\textdegree C of iron powder (Alfa Aesar 99.9+\%) and ground arsenic lumps (Alfa Aesar 99.9999\%) in a rotating silica ampoule as described in reference \onlinecite{Ran2014Thesis}. 

\begin{figure*}
	\includegraphics[width=\textwidth]{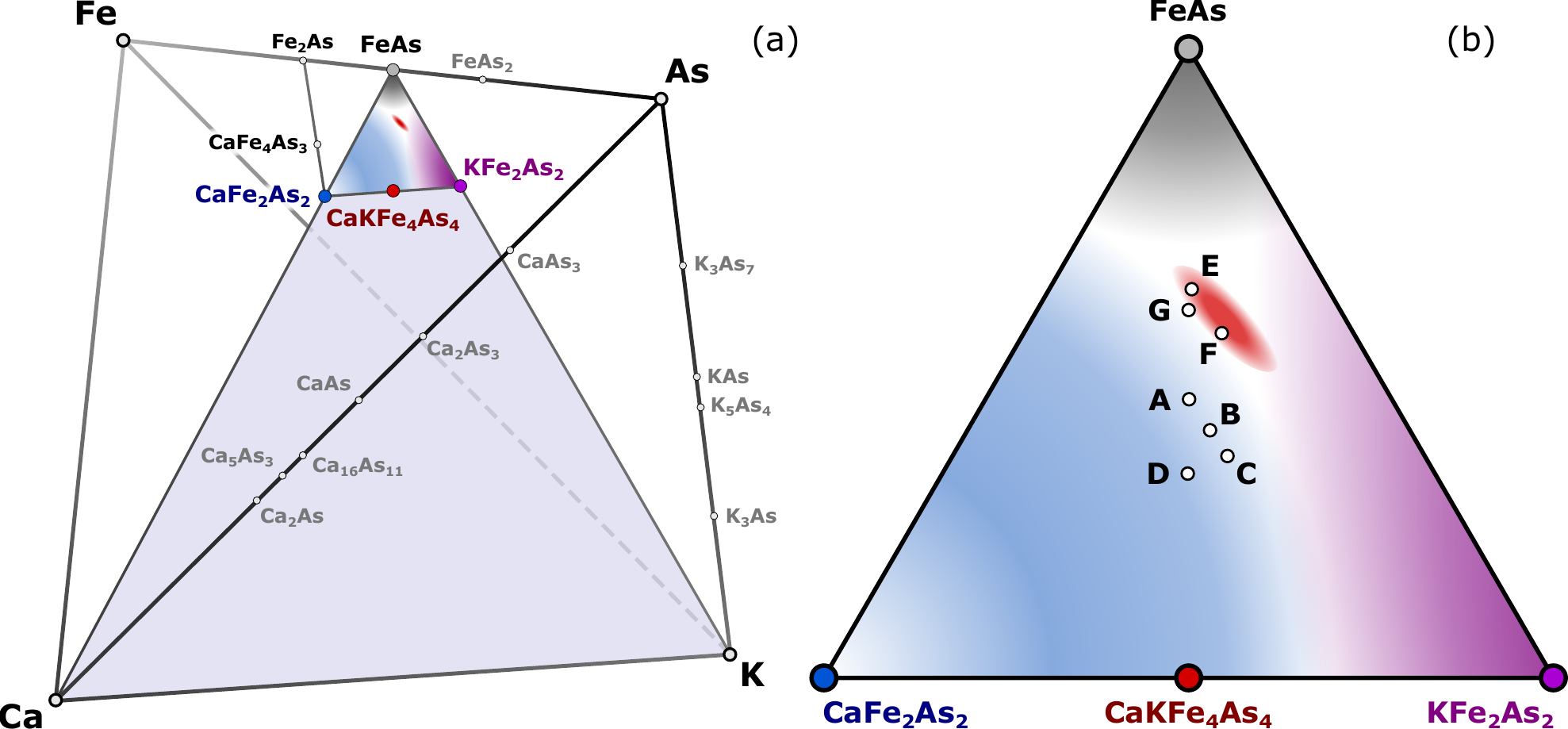}%

	\caption{(a) The 3-dimensional quaternary Ca-K-Fe-As phase diagram. The shaded blue-gray plane represents the compositions with equal iron and arsenic fractions. The red ellipse below FeAs represents the primary solidification region (primary phase field) of CaKFe$_{4}$As$_{4}$ (Compositions where pure single crystals can be grown) determined through optimization. (b) An enlarged pseudo-ternary phase diagram of composition near FeAs. The proposed primary solidification regions of each phase are shaded. Blue is CaFe$_{2}$As$_{2}$, gray is FeAs, red is CaKFe$_{4}$As$_{4}$ and, purple is KFe$_{2}$As$_{2}$. The labeled points are projections of the batch compositions in table~\ref{Compositions} onto the composition plane. There is some evidence that CaFe$_{2}$As$_{2}$ does not melt congruently and therefore would not be the primary phase from a liquid of its own composition. The extent of the KFe$_{2}$As$_{2}$ region is speculative.
		\label{QuaternaryTernary}}
\end{figure*}

The fritted alumina crucible set was assembled inside a 12.7\,mm OD tantalum tube which was sealed in a custom arc-welding chamber.\cite{Kong2015K2Cr3As3} The tantalum tube was sealed in a fused silica ampoule back-filled with approximately 1/5\,atm argon to protect the tantalum from oxidation at high temperatures. Figure~\ref{ampoule_profile}a presents a schematic of the complete ampoule assembly. The tantalum tube protects the silica from reactive potassium vapor and the alumina crucible protects the tantalum from attack by the solution. A growth attempt without the alumina crucible set proved unsuccessful as the quaternary liquid attacked the tantalum, leaked out, and damaged the silica ampoule. Precautions were taken to ensure that any failure would not allow arsenic to escape because of the toxicity of arsenic vapor and compounds (see description of furnace below).

The compositions within the quaternary Ca-K-Fe-As system can be represented as locations within a 3-dimensional tetrahedron shown in Fig.~\ref{QuaternaryTernary}a.\cite{Bergeron_IntroPhaseDiagrams} This is difficult to visualize and display. However, the iron-arsenic ratio is fixed by using Fe$_{0.512}$As$_{0.488}$ powder as a reactant. Compositions are projected onto the Fe\,:\,As\,=\,1\,:\,1 plane (bounded by Ca-K-FeAs and shaded in gray in Fig.~\ref{QuaternaryTernary}a) despite the slightly unequal Fe-As ratio. The region of this plane near FeAs is enlarged in Fig.~\ref{QuaternaryTernary}b. The compositions discussed below are labeled on this pseudo-ternary and summarized in table~\ref{Compositions}. 

%\begin{figure}
%	\includegraphics[width=\columnwidth]{KCaFe4As4_Ternary+liquidus3}%
%	\caption{A pseudo-ternary phase diagram of the system with the proposed primary solidification regions of each phase shaded. Blue is CaFe$_{2}$As$_{2}$, gray is FeAs, red is CaKFe$_{4}$As$_{4}$ and, purple is KFe$_{2}$As$_{2}$. The labeled points are projections of the batch compositions in table~\ref{Compositions} onto the composition plane. There is some evidence that CaFe$_{2}$As$_{2}$ does not melt congruently. Therefore would not be the primary phase from a liquid of its own composition. The extent of the KFe$_{2}$As$_{2}$ region is speculative.
%		\label{ternary}}
%\end{figure}

\begin{table}
	\caption{\label{Compositions}Selected Batched Compositions}
	\begin{ruledtabular}
		\begin{tabular}{c|c|c|c|c}
			\multirow{2}{*}{Composition} & \multicolumn{3}{c|}{Molar Ratios} & \multirow{2}{*}{Batches}\\
			& K & Ca & Fe$_{0.512}$As$_{0.488}$ & \\
			\hline
			A & 1 & 1 & 16 & A1, A2\\ %TU956, TU984
			B & 1.2 & 1 & 16 & B1\\ %TU964
			C & 1.4 & 1 & 16 & C1\\ %TU965
			D & 1 & 1 & 12.7 & D1\\ %TU961
			E & 1 & 1 & 24 & E1, E2, E3\\ %TU998, TU999, ML001
			F & 1.2 & 0.8 & 20 & F1, F2 \\ %ML019, ML292
			G & 1 & 1 & 22 & G1 \\ %ML017
		\end{tabular}
	\end{ruledtabular}
\end{table}

\subsection{Furnace Schedule}
The ampoule-crucible assemblies were processed in Lindburg Blue M, 1500\textdegree C box furnaces with silicon carbide heating elements. The furnaces are enclosed within fume hoods which mitigate the hazards associated with arsenic escaping at high temperatures. The furnace schedules evolved throughout the optimization process. All growths were heated over 1\,hour to 650\textdegree C and held there for 3\,hours to allow any free arsenic to react before ramping over 2\,hours to 1180\textdegree C. This temperature was held for 5\,hours to facilitate the formation of a homogeneous solution. 1180\textdegree C was chosen because it exceeded the melting temperature of FeAs, 1030\textdegree C,\cite{As-FePhaseDiagram} but is below about 1200\textdegree C where amorphous silica tends to deform.\cite{Canfield1992_GrowthMetallicFluxes} The furnace was then slowly cooled to induce crystallization.

After this slow cool, the furnace was held at the "spin temperature" until the ampoule was removed, inverted into a modified centrifuge, and spun.\cite{Canfield1992_GrowthMetallicFluxes,Canfield2001_HighTempSolutionGrowth} When successful, this decanting procedure removes any liquid to the other side of the fritted alumina crucible set and leaves clean crystals behind. The crucible set has a strainer (called a frit) that catches the crystals but allows the liquid to pass through.\cite{Canfield2016FritDiskCrucibles}

After the ampoule cooled to room temperature, the assembly was opened and examined in an argon or nitrogen filled glove-box. Whereas pure CaKFe$_{4}$As$_{4}$ crystals are visibly stable in air for months, the solidified liquid is air sensitive. 

The individual pieces of the crucible assembly were weighed before and after the growth in an attempt to estimate the mass of the products that remained after decanting. In cases where the liquid was cleanly separated from single phase crystals, these masses provided valuable information about the fractions of equilibrium phases when the assembly was removed from the furnace. The liquid composition at the time of decanting can be estimated by subtracting the quantity of the elements in the crystals from the amount initially batched. This composition defines a point on the compound's liquidus which can guide future growth compositions and refine the temperature ranges of crystal growth. Even when multiple solid phases are present (for example CaFe$_{2}$As$_{2}$ and CaKFe$_{4}$As$_{4}$) the range of possible liquid composition can be constrained to lie between the compositions estimated assuming either compound had been precipitated alone.

CaKFe$_{4}$As$_{4}$ forms as gray, metallic, 0.1-1\,mm thick plates that are, unfortunately, visually indistinguishable from CaFe$_{2}$As$_{2}$ or KFe$_{2}$As$_{2}$. The plates of 1144 are soft and exfoliate apart in the (001) into thin, slightly flexible sheets very similar to mica and tend to fracture along (100). The crystals often grow until they impinge on the liquid surface, another plate, or the crucible wall. This situation can lead to trapped volumes of liquid which are difficult to remove, even by centrifuging. Free edges of the plates are rounded and show no evidence of faceting other than the (001).

\subsection{Measurements}
The soft, micaceous nature of the CaKFe$_{4}$As$_{4}$ plate-like crystals leads them to smear in a mortar and pestle instead of grinding, similar to CaFe$_{2}$As$_{2}$.\cite{Ni2008TransnCaFe2As2} This makes it difficult to prepare a powder sample suitable for x-ray diffraction. Instead we affixed a single crystal to the sample holder with vacuum grease (Dow Corning High Vacuum Grease) in a Rigaku Miniflex II diffractometer (with a Cu tube and monochromator) and used the method described in Jesche et al.\cite{Jesche2016CrystalXRD} to obtain (00$\ell$) lines from the crystals and powder patterns from any polycrystalline surface impurities. A disadvantage of this method is that the laboratory diffractometer's x-rays only probe the crystal's surface and therefore may not accurately represent the phase-fractions of bulk.

\begin{figure}
	\includegraphics[width=\columnwidth]{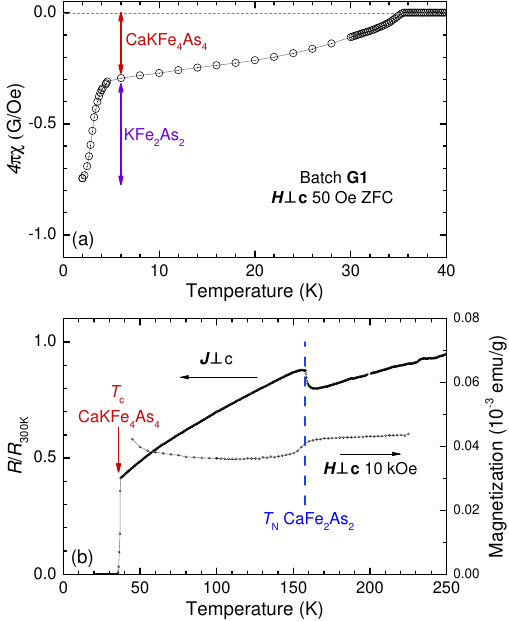}%
	\caption{Signatures of the iron arsenide phases in (a) zero-field-cooled (ZFC) low field magnetization, (b) resistance and high-field magnetization vs. temperature measurements of a crystal from batch \textbf{G1}. Below 35\,K, CaKFe$_{4}$As$_{4}$ contributes to the diamagnetic response in panel (a) and zero resistance  in (b). The diamagnetic response below 4\,K is attributed to KFe$_{2}$As$_{2}$. Non-superconducting CaFe$_{2}$As$_{2}$ generates a jump in resistance and magnetization in panel (b) at its magnetic-structural transition near 160\,K.
		\label{3Phase}}
\end{figure}

\begin{figure*}
	\includegraphics[width=\textwidth]{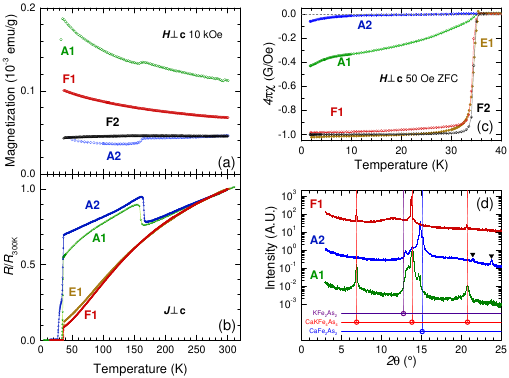}%
	\caption{(a) High temperature magnetization, (b) resistance, (c) zero-field-cooled (ZFC) low temperature magnetization, and (d) x-ray diffraction (Cu K$_\alpha$ source) of (00$\ell$) oriented single crystals used to identify arsenide phases present in described batches. Below 35\,K, CaKFe$_{4}$As$_{4}$ contributes to the diamagnetic response in panel (c) and zero resistance  in (b). The diamagnetic response of batches \textbf{A1} and \textbf{A2} in (c) below 4\,K is attributed to the superconducting shielding of KFe$_{2}$As$_{2}$. Non-superconducting CaFe$_{2}$As$_{2}$ in batches \textbf{A1} an \textbf{A2} generates a jump in magnetization in panel (a) and resistance in (b) at its magnetic-structural transition near 160\,K. The Curie-Weiss-like components of magnetization in figure (a) (\textbf{A1} and \textbf{F1}) are attributed to FeAs, Fe$_{2}$As or their corrosion products. The plots in (d) are shifted for clarity and the peaks marked with triangles do not correspond to (00$\ell$) lines an of the 3 labeled phases. They may be due to misoriented grains or a binary iron arsenide phases.
		\label{Story}}
\end{figure*}

The similar, in-plane, lattice parameters of CaKFe$_{4}$As$_{4}$ (\textit{a}\,=\,0.3866\,nm)\cite{Iyo2016AeAFe4As4}, CaFe$_{2}$As$_{2}$ (\textit{a}\,=\,0.3912\,nm)\cite{Ni2008TransnCaFe2As2}, and KFe$_{2}$As$_{2}$ (\textit{a}\,=\,0.38414\,nm)\cite{Sasmal2008A1-xSrxFe2As2} and their close structural relationship leads to co-aligned, epitaxial growth of these phases on each other. As a consequence, a plate that appears to be a single crystal may actually contain two or three phases. This concern was confirmed by high energy synchrotron x-ray diffraction measurements in transmission mode which show aligned peaks from all three phases.\cite{Meier2016CaKFe4As4} This situation lead us to utilize and, ultimately, emphasize magnetization and resistance measurements to reveal the presence of the impurity CaFe$_{2}$As$_{2}$ and KFe$_{2}$As$_{2}$ phases in individual samples. The x-ray diffraction pattern of a crystal from batch \textbf{A1} in Fig.~\ref{Story}d (below) shows the presence of all three phases within, what appears to be, a single crystal.

A Quantum Design Magnetic Property Measurement System (MPMS) was used to measure the temperature dependence of magnetization and resistance. Crystals selected for magnetization measurements were relatively clean of surface impurities and positioned so that the applied field was in the plane of the plate (perpendicular to the crystallographic c-axis). In this geometry, the demagnetization factor, $D\approx 0$. In order to reveal superconducting phases, samples were zero-field-cooled (ZFC) to 1.8\,K at which point, a field of 50 Oe was applied and the magnetization was measured on warming. Note that this field is well below the $H_{c1}$$=$1000\,Oe of CaKFe$_{4}$As$_{4}$.\cite{Meier2016CaKFe4As4} The density of CaKFe$_{4}$As$_{4}$, determined from lattice parameters, 5.22\,g/cm$^{3}$,\cite{Iyo2016AeAFe4As4} was used to calculate the volumetric magnetic susceptibility, $\chi$, in order to estimate superconducting volume fraction. Note that even in multiphase samples this is a reasonable approximation because the densities of CaFe$_{2}$As$_{2}$ and KFe$_{2}$As$_{2}$ are within 10\% of CaKFe$_{4}$As$_{4}$.\cite{Ni2008TransnCaFe2As2,Sasmal2008A1-xSrxFe2As2} If the effect of the demagnetization field is negligible, as we would expect for field in the plane of a plate, $4\pi\chi = -1$ would be full superconducting shielding. In addition, the magnetic susceptibility at 10\,kOe was measured up to 300\,K. Four-probe, in plane ($\textit{\textbf{J}}$$\perp$$c$) resistance was also measured in a Quantum Design MPMS using an AC resistance bridge (Linear Research Model LR-700) with platinum leads bonded to the sample with silver paint.

Figure~\ref{3Phase} presents magnetization and resistance data taken on a crystal from batch \textbf{G1} that contains intergrown CaKFe$_{4}$As$_{4}$, CaFe$_{2}$As$_{2}$, and KFe$_{2}$As$_{2}$ to illustrate their individual signatures. Low magnetic field, magnetization vs. temperature scans are the most sensitive test for superconducting CaKFe$_{4}$As$_{4}$ and the KFe$_{2}$As$_{2}$ impurity phase, which have superconducting transition temperatures of 35\,K and 3.8\,K respectively.\cite{Meier2016CaKFe4As4,Rotter2008_Ba1-xKxFe2As2WholeRange} The magnetic susceptibility data in Fig.~\ref{3Phase}a displays the superconducting shielding of these two superconducting phases. The reduction of a sample's diamagnetic response on warming through about 4\,K is due to the superconducting transition of KFe$_{2}$As$_{2}$ and provides an estimate of its volume fraction, possibly 50\,vol\%. The other contribution to the diamagnetic signal disappears upon heating through 35\,K and suggests a moderate volume, about 30\,vol\%, of CaKFe$_{4}$As$_{4}$. The remaining 20\,vol\% is attributed to non-superconducting phases. 

CaFe$_{2}$As$_{2}$ does not superconduct at ambient pressure but undergos a paramagnetic, tetragonal to orthorhombic, antiferromagnetic phase transition at 170\,K.\cite{Ni2008TransnCaFe2As2} If this phase comprises a sufficient volume fraction, the transition produces a jump in magnetization vs. temperature, demonstrated in Fig.~\ref{3Phase}b at about 160\,K. Strain or small substitutions of K for Ca could explain the disagreement with the previously reported transition temperature of pure CaFe$_{2}$As$_{2}$. The Curie-Weiss-like upturn of magnetization at low temperatures in Fig.~\ref{3Phase}b is associated with the binary iron arsenides FeAs and Fe$_{2}$As\cite{Onnerud1995_TetFe2-xCoxAs}, or their corrosion products.

Resistance vs. temperature measurements are very sensitive to any superconducting phases as voltage drops to zero as soon as a superconducting path forms. In Fig.~\ref{3Phase}b resistance is zero below 35\,K indicating the presence of superconducting CaKFe$_{4}$As$_{4}$. This figure also demonstrates a sharp jump in resistance at the magnetic and structural transition of CaFe$_{2}$As$_{2}$. This jump is the dominant signature of CaFe$_{2}$As$_{2}$ and persists even when the corresponding feature in magnetization can not be readily resolved.

\section{Optimization of Crystal Growth}

Our first attempt to grow CaKFe$_{4}$As$_{4}$, batch \textbf{A1}, was based on our previous growths of CaFe$_{2}$As$_{2}$\cite{Ran2012CaFe2-2xCo2x2As2_Annealing} and used composition \textbf{A} on Fig.~\ref{QuaternaryTernary}b and Table~\ref{Compositions} (K\,:\,Ca\,:\,Fe$_{0.518}$As$_{0.488}$\,=\,1\,:\,1\,:\,16). In these early growths, the temperature was reduced from 1180\textdegree C directly to the final temperature, in this case 960\textdegree C, over 28\,hours. The liquid was completely decanted yielding 5-10\,mm metallic, plate-like crystals free-standing in the crucible. The initially white alumina crucibles were observed to turn black, not just where the liquid had wetted the interior wall, but all over. We suspected that this was a result of the attack by potassium vapor inside the tantalum tube. 

Resistance vs. temperature and magnetization vs. temperature measurements on the plates (\textbf{A1} in Fig.~\ref{Story}b and c) indicate superconductivity at about 35\,K. This is similar to the report of $T_{\mathrm{c}}$=\,33.1\,K for CaKFe$_{4}$As$_{4}$ in Iyo et al.\cite{Iyo2016AeAFe4As4} In addition there is also a jump in resistance that corresponds to the structural/magnetic transition temperature of CaFe$_{2}$As$_{2}$. ZFC magnetization vs. temperature (\textbf{A1} in Fig.~\ref{Story}c) suggests a moderate, but incomplete, superconducting shielding of approximately 35\,vol\% below 35\,K. There also may be some small quantity of KFe$_{2}$As$_{2}$ producing a larger diamagnetic moment below 4\,K. X-ray diffraction data of from a crystal from batch \textbf{A1} in Fig.~\ref{Story}d further supports the presence of CaKFe$_{4}$As$_{4}$, CaFe$_{2}$As$_{2}$, and KFe$_{2}$As$_{2}$ in a single plate.

Other compositions (\textbf{B}, \textbf{C} and \textbf{D} on Fig.~\ref{QuaternaryTernary}b) decanted at 960\textdegree C yielded similar multiphase products. Finding significant fractions of CaKFe$_{4}$As$_{4}$ and CaFe$_{2}$As$_{2}$ for all these compositions suggested that both phases had crystallized out of solution before the liquid was decanted at 960\textdegree C. It is likely that one of the compounds crystallized out of solution at a higher temperature before the other precipitated epitaxially on top. 

To test this hypothesis, we batched the composition \textbf{A} again and cooled more slowly from the 1180\textdegree C hold to 1000\textdegree C (rather than 960\textdegree C) over 100\,hours. The resulting crystals exhibit a broad drop to zero resistance in Fig.~\ref{Story}b (\textbf{A2}) as well as the jump associated with the structural transition of CaFe$_{2}$As$_{2}$. Low temperature magnetization measurements show almost no diamagnetic shielding (\textbf{A2} in Fig.~\ref{Story}c) and indicate a very small fraction of CaKFe$_{4}$As$_{4}$ (on the order of 1\,vol\%) and a minor amount of KFe$_{2}$As$_{2}$. Magnetization results at higher temperatures (\textbf{A2} in Fig.~\ref{Story}a) show a discontinuity at the CaFe$_{2}$As$_{2}$ structural/magnetic transition. These data suggest that the crystals consist primarily of CaFe$_{2}$As$_{2}$ without a significant fraction of CaKFe$_{4}$As$_{4}$. X-ray diffraction supports this conclusion and shows strong CaFe$_{2}$As$_{2}$ peaks in Fig.~\ref{Story}d. A similar result was found for composition \textbf{B} decanted at 1000\textdegree C.

Obtaining fairly clean CaFe$_{2}$As$_{2}$ at 1000\textdegree C and mixed CaKFe$_{4}$As$_{4}$ and CaFe$_{2}$As$_{2}$ at 960\textdegree C clearly indicates the order of solidification. The primary phase, CaFe$_{2}$As$_{2}$, crystallizes out of solution first on cooling for compositions \textbf{A} and \textbf{B} and CaKFe$_{4}$As$_{4}$ follows starting somewhere between 1000\textdegree C and 960\textdegree C. 

Moving the batched composition away from CaFe$_{2}$As$_{2}$ should suppress its formation. The temperature at which crystallization would start was unknown, therefore we assembled three batches with composition \textbf{E} (K\,:\,Ca\,:\,Fe$_{0.518}$As$_{0.488}$\,=\,1\,:\,1\,:\,24). All three were placed in the same furnace and rapidly cooled from 1180\textdegree C to 1100\textdegree C over 1\,hr and then slowly cooled. At each of the following temperatures: 960\textdegree C, 940\textdegree C, and 920\textdegree C, one batch was removed and decanted. Cooling quickly to 1100\textdegree C not only reduces the time wasted cooling a homogeneous liquid, but also reduced the attack of the alumina crucibles by potassium vapor by spending less time at the highest temperatures. 

Batch \textbf{E1}, decanted at 920\textdegree C, was the only batch that had formed crystals. In fact, the material had almost entirely solidified over the final 20\textdegree C. The resulting material consisted of metallic plates embedded in a matrix of solidified gray, sub-metallic polycrystals with a few small cavities. These cavities were likely filled with the final liquid present when the assembly was decanted. The narrow temperature range of solidification and small residual liquid fraction suggest that composition \textbf{E} is close to that of a eutectic near 920\textdegree C. A eutectic liquid solidifies into two or more solid phases as it is cooled through the eutectic temperature. Compositions nearby the eutectic will solidify over a narrow temperature range just above this temperature.\cite{Bergeron_IntroPhaseDiagrams} 

Resistance and magnetization vs. temperature measurements (\textbf{E1} Fig.~\ref{Story}b and c) were carried out on sheets of the crystalline plates exfoliated out of the solidified matrix. The sharp transition from full superconducting magnetic shielding to normal state on heating through 35\,K suggest that they are crystals of nearly phase pure CaKFe$_{4}$As$_{4}$. Powder x-ray diffraction measurements on the sub-metallic polycrystals show that they are primarily the binary iron arsenides FeAs and Fe$_{2}$As. This assemblage of phases and the narrow temperature range over which they solidified support the claim that composition \textbf{E} is close in composition to the CaKFe$_{4}$As$_{4}$-FeAs eutectic near 920\textdegree C.

Although we have obtained fairly clean CaKFe$_{4}$As$_{4}$ from composition \textbf{E}, the narrow temperature range over which complete solidification occurs makes it difficult to obtain free-standing plates. To further optimize the crystal growth we can take some guidance from thermodynamics. If we consider the projection of composition \textbf{E} onto the Fe\,:\,As\,=\,1\,:\,1 plane in Fig.~\ref{QuaternaryTernary}b, it lies on the a line between CaKFe$_{4}$As$_{4}$ and FeAs. In fact, this is the Alkemade line that connects two phases.\cite{Levin1986PhaseDiagramsForCeramists} Alkemade's theorem\cite{Alkemade1893AlkemadeTheorem} suggests that the maximum temperature at which these two phases are in thermodynamic equilibrium with a liquid occurs when the liquid composition lies on the Alkemade line between them, or its extrapolation.\cite{Levin1986PhaseDiagramsForCeramists,Malakhov2004AlkemadeTheorem} This means that batches with compositions farther from the line between CaKFe$_{4}$As$_{4}$ and FeAs should completely solidify at increasingly lower temperatures. This will also provide a larger temperature range over which the only equilibrium phases are CaKFe$_{4}$As$_{4}$ and liquid, allowing us to decant the liquid and obtain free-standing crystals.

Starting from an assumed binary eutectic near composition \textbf{E} on Fig.~\ref{QuaternaryTernary}b, we would expect a finite liquid fraction to exist below the eutectic temperature for compositions to the left and right of \textbf{E}. If we move to the left, by replacing some potassium with calcium, the composition moves closer to CaFe$_{2}$As$_{2}$. As we already believe this phase to be less soluble than CaKFe$_{4}$As$_{4}$, it would likely crystallize first and might prevent us from obtaining pure crystals of the latter. If we move to the right, replacing calcium with potassium, we may be able to obtain our target phase. Since this composition is now farther from CaFe$_{2}$As$_{2}$ we may still be able to avoid precipitating the CaFe$_{2}$As$_{2}$ phase even at lower concentrations of iron and arsenic. Based on these arguments we choose to batch composition \textbf{F}.

Batch \textbf{F1}, with composition \textbf{F} (K\,:\,Ca\,:\,Fe$_{0.518}$As$_{0.488}$\,=\,1.2\,:\,0.8\,:\,20), was cooled quickly from 1180\textdegree C to 1050\textdegree C over 2\,hours then slowly cooled to 930\textdegree C over 30.5\,hours and decanted. The start temperature of the slow cool, 1050\textdegree C, was selected based on the results from previously tested compositions. The estimated liquid composition of batch \textbf{A2} at 1000\textdegree C was near composition \textbf{F}. This gave us some confidence that composition \textbf{F} would still be a homogeneous liquid at 1050\textdegree C. This adjustment reduced the time at the highest temperatures where K likely reacts more rapidly with the alumina crucible. The final temperature was increased to 930\textdegree C in order to avoid the dramatic solidification observed at 920\textdegree C described above. Reducing the temperature range for slow cooling also means that more time can be spent in the temperature range where the crystals are growing.

\begin{figure}
	\includegraphics[width=0.6 \columnwidth]{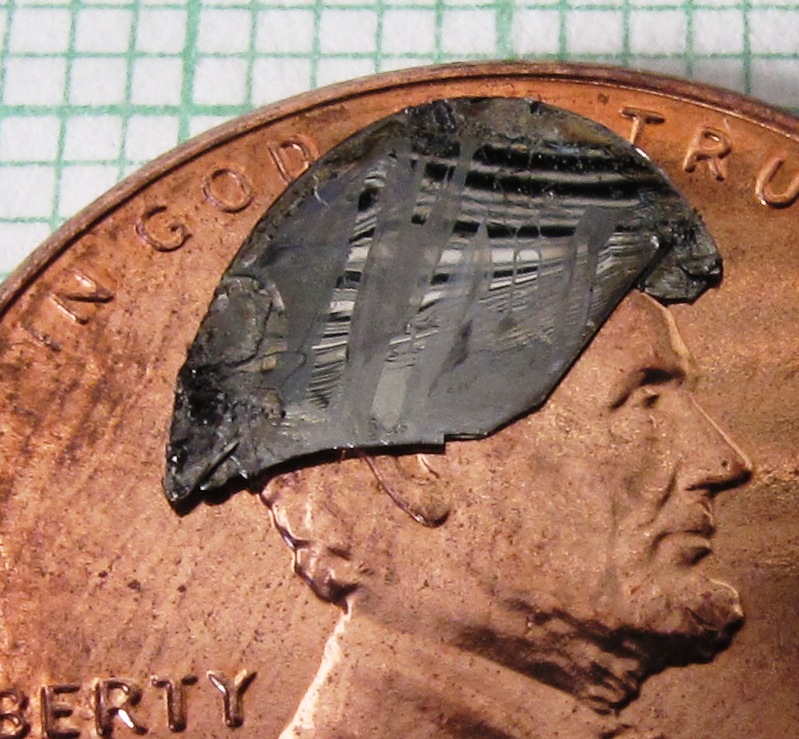}%
	
	\caption{Large crystal of CaKFe$_{4}$As$_{4}$ from batch \textbf{F1} on a penny for scale. A 1\,mm grid is visible at the top.\cite{CrystalPicNote}}
	\label{TrumpCrystal}

\end{figure}

Batch \textbf{F1} yielded large (up to 5x10\,mm$^{2}$) free-standing metallic plates. Some of these crystals were limited in size by the crucible dimensions. This gives them round or elliptical edges as exhibited by the crystal in Fig.~\ref{TrumpCrystal}. The faces of the plates were determined to be parallel (001) with or without curved terraces. The crystals also cleave along the (001). No other facets were observed and freely growing plate edges are often curved or scalloped.

The surfaces of the plates often had a few small (0.1-0.3\,mm) clusters of gray sub-metallic polycrystals which are likely solidified droplets of liquid. When this polycrystalline material is exposed to air it appears to boil, fizz and turn black. Possibly this material contains of a hygroscopic potassium compound that reacts with moisture in air. 

The plates from \textbf{F1} appear to be relatively pure CaKFe$_{4}$As$_{4}$. They exhibit near perfect superconducting shielding (\textbf{F1} in Fig.~\ref{Story}c) and a sharp transition from the superconducting to normal state. In addition, the relationship between $T_{\mathrm{c}}$ and the specific heat jump\cite{Meier2016CaKFe4As4} consistent are with BNC scaling\cite{Sergey09BNC} for Fe-based superconductors. The presence of binary iron arsenides (or their corrosion products) on the crystal in Fig.~\ref{TrumpCrystal} are evident in the Curie Weiss divergence of its normal state magnetic susceptibility (\textbf{F1} in Fig.~\ref{Story}a). There are no features in resistance vs. temperature (\textbf{F1} Fig.~\ref{Story}b) other than superconductivity. In addition, the residual resistance ratio, $R_{300\,K}/R_{T_{c}}$, has increased dramatically from about 2 in early batches to 15.

Considering all of these physical properties, in addition to x-ray diffraction (Fig.~\ref{Story}d), the results indicate that there is little to no CaFe$_{2}$As$_{2}$ or KFe$_{2}$As$_{2}$ in these crystals from \textbf{F1}. This suggests that CaKFe$_{4}$As$_{4}$ crystallizes out of the liquid first for composition \textbf{F}.

Crystals were improved in subsequent growths by further narrowing of the slow cooling range by starting at 990\textdegree C instead of 1050\textdegree C and cleaving off surfaces of the crystals that had polycrystalline second phases, such as binary iron arsenides. Cleaving off undesirable phases is only possible now that CaFe$_{2}$As$_{2}$ and KFe$_{2}$As$_{2}$ are not inter-grown with the crystals. The magnetic susceptibility of sample from batch \textbf{F2} shows the sharpest loss of diamagnetic shielding at $T_{\mathrm{c}}$ in Fig.~\ref{Story}c. The same sample also demonstrates the intrinsic normal state magnetic susceptibility (\textbf{F2} Fig.~\ref{Story}a) without the Curie Weiss contribution associated with binary iron arsenides. Note the similarity in the magnitude of paramagnetic susceptibility of CaKFe$_{4}$As$_{4}$ (\textbf{F2}) and CaFe$_{2}$As$_{2}$ (\textbf{A2}).

\section{DISCUSSION}
\begin{figure}
	\includegraphics[width=\columnwidth]{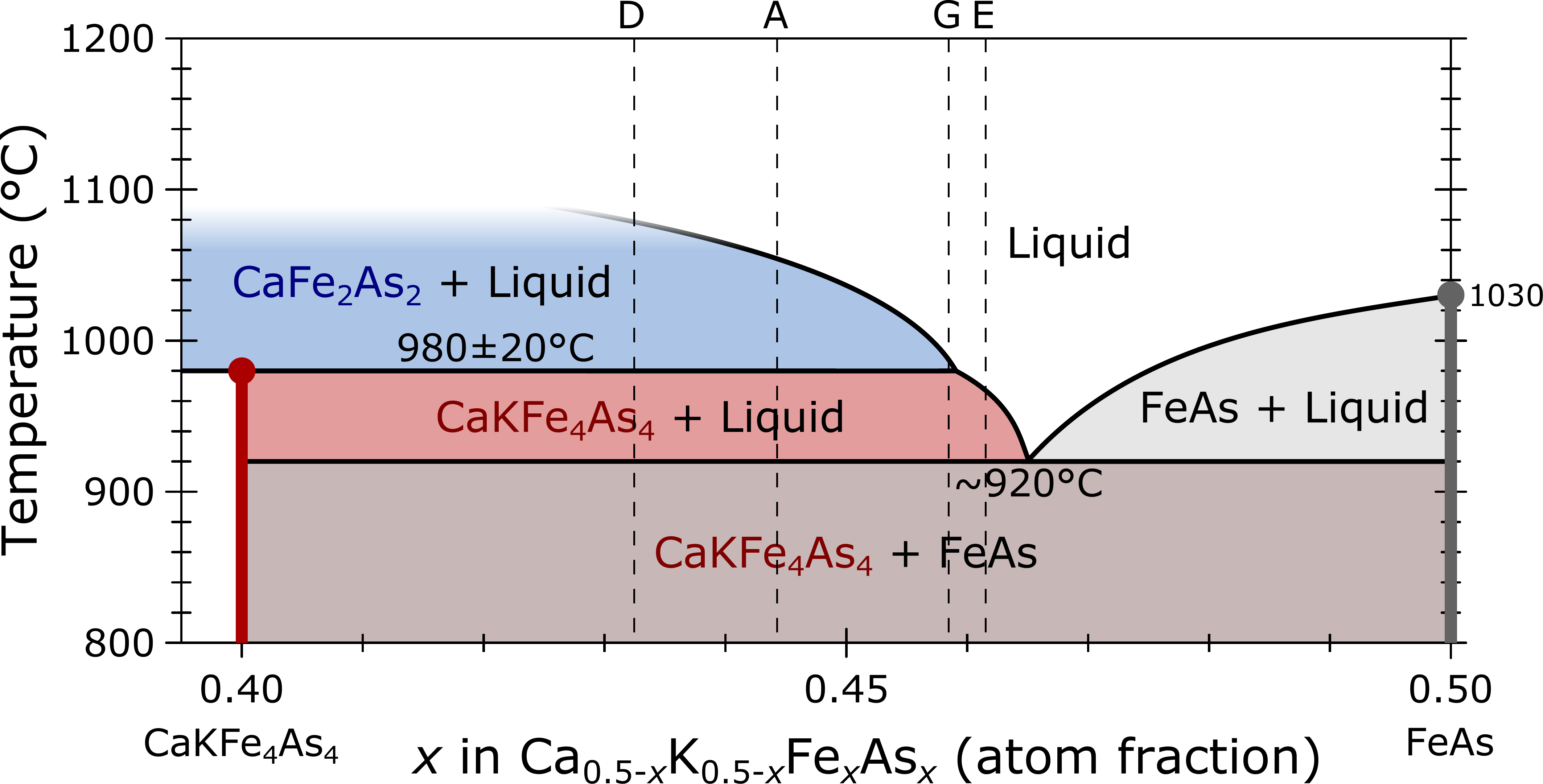}%
	\caption{A schematic of the proposed vertical section from CaKFe$_{4}$As$_{4}$ to FeAs based on batches described in this study. Each colored area is labeled with the phases present within its composition-temperature range. The vertical dashed lines indicate the locations of compositions \textbf{D}, \textbf{A}, \textbf{G} and, \textbf{E} in Fig.~\ref{QuaternaryTernary}b and table~\ref{Compositions}. The optimized composition \textbf{F} does not have equal fractions of Ca and K and therefore is not displayed in this section but it would lie near $x$\,$=$\,$0.455$. CaKFe$_{4}$As$_{4}$ is indicated to melt incongruently via a peritectic reaction to CaFe$_{2}$As$_{2}$ and liquid. This is indicated by the termination of its red phase line at $x$\,$=$\,$0.4$ at 980\textdegree C. 
		\label{pseudobinary}}
\end{figure}

The described series of growths give a glimpse of the quaternary phase diagram in the composition region near FeAs-CaFe$_{2}$As$_{2}$-KFe$_{2}$As$_{2}$. As we chased the CaKFe$_{4}$As$_{4}$ phase we delineated many of the primary solidification regions (primary phase fields). We present our interpretation of these areas schematically as colored regions in Fig.~\ref{QuaternaryTernary}b. In addition, an inferred vertical section from CaKFe$_{4}$As$_{4}$ (Ca$_{0.1}$K$_{0.1}$Fe$_{0.4}$As$_{0.4}$) to FeAs (Fe$_{0.5}$As$_{0.5}$) is sketched out in Fig.~\ref{pseudobinary}.

Plates obtained from compositions \textbf{A}-\textbf{D} decanted at 960\textdegree C were determined to be mixed phase CaFe$_{2}$As$_{2}$ and CaKFe$_{4}$As$_{4}$. But, decanting compositions \textbf{A} and \textbf{B} at 1000\textdegree C yielded primarily CaFe$_{2}$As$_{2}$. This result clearly demonstrates that CaFe$_{2}$As$_{2}$ crystallizes first on cooling for liquids in this composition range. It is followed by CaKFe$_{4}$As$_{4}$ which starts to crystallize between 960\textdegree C and 1000\textdegree C. 

This is shown in Fig.~\ref{pseudobinary} as a CaFe$_{2}$As$_{2}$\,+\,Liquid region changing to a CaKFe$_{4}$As$_{4}$\,+\,Liquid equilibrium on cooling through about 980\textdegree C. This situation is consistent with a peritectic reaction. As the system is cooled through the peritectic temperature the liquid reacts with an existing phase (CaFe$_{2}$As$_{2}$) to form a new solid phase (CaKFe$_{4}$As$_{4}$.)\cite{Bergeron_IntroPhaseDiagrams} There is likely some temperature-composition region within the range presented in Fig.~\ref{pseudobinary} where both solid phases and the liquid are stable. However, we have omitted this for simplicity and because we do not have any indications of its extent. 

It should be kept in mind that solidification is often not an equilibrium process and peritectic-like reactions often do not progress to completion.\cite{Bergeron_IntroPhaseDiagrams} This tendency for non-equilibrium phase assemblages would be exacerbated by the epitaxial growth of CaKFe$_{4}$As$_{4}$ on CaFe$_{2}$As$_{2}$. The diffusion of Ca$^{2+}$ and K$^{+}$ ions through the encasing CaKFe$_{4}$As$_{4}$ layer is probably slower than through the liquid. As a result, the rate of the peritectic reaction would be impeded by the growing shell of its product. It is worth noting that Fig.~\ref{pseudobinary} is a vertical section and that the equilibrium phases do not have to have compositions within the figure.\cite{Bergeron_IntroPhaseDiagrams} For example, in the blue CaFe$_{2}$As$_{2}$\,+\,Liquid field, neither CaFe$_{2}$As$_{2}$ nor the equilibrium liquid have a calcium to potassium ratio of one. This means that these phases do not lie on the plane of Fig.~\ref{pseudobinary}. 

Non-equilibrium solidification may explain why crystals from batch \textbf{G1} (composition \textbf{G} in Fig.~\ref{QuaternaryTernary}b and \ref{pseudobinary}) contain CaKFe$_{4}$As$_{4}$, CaFe$_{2}$As$_{2}$ and KFe$_{2}$As$_{2}$ as indicated in Fig.~\ref{3Phase}. If CaFe$_{2}$As$_{2}$ solidifies first (as we suggested in Fig.~\ref{pseudobinary}) then some KFe$_{2}$As$_{2}$ could precipitate before all the CaFe$_{2}$As$_{2}$ reacts with the liquid to form CaKFe$_{4}$As$_{4}$

The relatively clean CaFe$_{2}$As$_{2}$ crystals obtained from batches \textbf{A2} and \textbf{B2} suggest that CaFe$_{2}$As$_{2}$ is stable above at least 1000\textdegree C. Yi et al. suggests that CaFe$_{2}$As$_{2}$ decomposes or melts between 900\textdegree C and 1200\textdegree C.\cite{Yi2011CaFe4As3} We propose that the primary solidification field of CaFe$_{2}$As$_{2}$, blue on in Fig.~\ref{QuaternaryTernary}b, extends from near the compound in the bottom left through compositions \textbf{A}-\textbf{D} and \textbf{G}. It is possible that CaFe$_{2}$As$_{2}$ decomposes into CaFe$_{4}$As$_{3}$ and other phases as proposed by Yi et al.\cite{Yi2011CaFe4As3} The composition of CaFe$_{4}$As$_{3}$ is plotted on Fig.~\ref{QuaternaryTernary}a lying between Fe$_{2}$As and CaFe$_{2}$As$_{2}$.

We suspect that CaFe$_{2}$As$_{2}$ is the primary phase at composition of CaKFe$_{4}$As$_{4}$ (Ca$_{0.1}$K$_{0.1}$Fe$_{0.4}$As$_{0.4}$) as well. This situation could occur if CaKFe$_{4}$As$_{4}$ melts incongruently via a peritectic reaction into CaFe$_{2}$As$_{2}$ and liquid. The temperature series at composition \textbf{E}, described above, and batch \textbf{F1} suggest that both lie within the primary solidification composition region of CaKFe$_{4}$As$_{4}$, the red field in Fig.~\ref{QuaternaryTernary}b. This is represented on the Fig.~\ref{pseudobinary} by the liquidus line of the 1144 phase marking the right-hand edge of the red CaKFe$_{4}$As$_{4}$\,+\,Liquid region. This supports our claim that CaKFe$_{4}$As$_{4}$ melts incongruently as it can only be crystallized alone from a liquid with a composition quite dissimilar from its own. The decomposition of CaKFe$_{4}$As$_{4}$ is depicted as an end to its vertical red phase line at $x=0.4$ in Fig.~\ref{pseudobinary} at about 980\textdegree C. The equilibria represented here also reflect the narrow temperature range that composition \textbf{E} is believed to have solidified over.

The eutectic inferred from composition \textbf{E} also represents the Ca-K-rich edge of the FeAs primary solidification area represented by gray in Fig.~\ref{QuaternaryTernary}b. As discussed above, Alkemade's theorem suggests that compositions near the eutectic will be liquid at lower temperatures farther from Ca:K = 1:1. This will lead to a wider temperature range over which only CaKFe$_{4}$As$_{4}$ and liquid are stable and single crystals of the former can be obtained by decanting.

On the potassium rich side of Fig.~\ref{QuaternaryTernary}b, there was little evidence obtained about the extent of the purple KFe$_{2}$As$_{2}$ primary solidification surface. Figure~\ref{Story}c indicates batches \textbf{A1} and \textbf{A2} both display small fractions of this phase and batch \textbf{G1} may have 50\,vol\% (Fig.~\ref{3Phase}a). CaFe$_{2}$As$_{2}$ is also present in all three of these cases and we believe it precipitated first. Crystallizing this phase depletes the melt in Ca which could lead to supersaturated KFe$_{2}$As$_{2}$. As mentioned above, the rate of the proposed peritectic reaction of CaFe$_{2}$As$_{2}$ and liquid to form CaKFe$_{4}$As$_{4}$ is diffusion limited. This non-equilibrium situation could explain the KFe$_{2}$As$_{2}$ observed for compositions where we do not believe it to be an equilibrium phase.

Sasmal et al. report obtaining KFe$_{2}$As$_{2}$ by solid state synthesis at 950\textdegree C.\cite{Sasmal2008A1-xSrxFe2As2} This provides a lower limit on its decomposition or melting temperature. In their supplemental materials, Zocco et al. state that "only samples of minor quality and small size were obtained" from crystal growths from Fe-As rich solutions.\cite{Zocco2013MultibandKFe2As2} This suggests that there is a small temperature window for KFe$_{2}$As$_{2}$ to grow before complete solidification. A likely situation is a eutectic between KFe$_{2}$As$_{2}$ and FeAs indicated by the boundary between the purple and gray areas in Fig.~\ref{QuaternaryTernary}b. This is probably why most single crystal studies of KFe$_{2}$As$_{2}$ obtained crystals from a KAs-rich solution.\cite{Zocco2013MultibandKFe2As2,Liu2014Ba1-xKxFe2As2_Growth,Kihou2010KFe2As2_KAsFlux,Terashima2009KFe2As2} Based on these considerations we believe that there is some primary solidification region of KFe$_{2}$As$_{2}$ present in the composition space depicted but, the extent shown is highly speculative.

\section{CONCLUSIONS}
The process described above demonstrates how we used a variety of techniques to inform the optimization of a growth protocol for CaKFe$_{4}$As$_{4}$. Resistance and magnetization measurements were frequently more valuable than x-ray diffraction in detecting CaFe$_{2}$As$_{2}$ and KFe$_{2}$As$_{2}$ impurities. These characterization techniques and Alkemade's theorem guided our choice of growth compositions and procedures. Estimation of liquid compositions at decanting temperature leveraging the Canfield Crucible Set guided the selection of growth compositions and more targeted growth temperature ramps. Taking full advantage of these diverse techniques expedited the discovery of successful growth parameters for single crystalline CaKFe$_{4}$As$_{4}$.

\begin{acknowledgements}
We would like to thank A. B\"{o}hmer for her assistance with preparation of this manuscript. In addition, A. Kreyssig and G. Drachuck for useful discussions and experimental assistance. This work was supported by the U.S. Department of Energy, Office of Basic Energy Science, Division of Materials Sciences and Engineering. The research was performed at the Ames Laboratory. Ames Laboratory is operated for the U.S. Department of Energy by Iowa State University under Contract No. DE-AC02-07CH11358. W. M. was supported by the Gordon and Betty Moore Foundation’s EPiQS Initiative through Grant GBMF4411.

\end{acknowledgements}

* Current Affiliation: Department of Chemistry, Princeton University, Princeton, New Jersey 08544

\bibliographystyle{apsrev4-1}
%\bibliography{KCaFe4As4_GrowthPaper_refs}

\begin{thebibliography}{35}
\bibitem [{Cry()}]{CrystalPicNote}%
	\BibitemOpen
	\href@noop {} {}\bibinfo {note} {Any resemblance to the hair or hue of any
		president, living or dead, is coincidental.}\BibitemShut {Stop}%
\end{thebibliography}
%

\end{document}